\documentclass[aps,prl, twocolumn,superscriptaddress]{revtex4-2}
\usepackage{amsmath,amssymb}
\usepackage{booktabs} 
\usepackage{graphicx}
\usepackage{xcolor}
\usepackage{array}
\usepackage{hyperref}
\usepackage{tikz}
\usetikzlibrary{positioning, fit, shapes}
\usepackage{pgfplots}

\begin{document}

\title{Rapid Graviton Thermalization through a Scalar--Gauss--Bonnet Coupling}

\author{Hongguang Zhang}
\email{zhanghg@bjtu.edu.cn}
\affiliation{Weihai Research Institute, Beijing Jiaotong University, Weihai 264401, China}

\author{Zhengyang Qiu}
\email{zhengyangqiu@bjtu.edu.cn}
\affiliation{Weihai Research Institute, Beijing Jiaotong University, Weihai 264401, China}

\author{Sirui Wang}
\email{siruiwang@bjtu.edu.cn}
\affiliation{Weihai Research Institute, Beijing Jiaotong University, Weihai 264401, China}

\author{Jiayi Wei}
\email{weijiayi@bjtu.edu.cn}
\affiliation{Weihai Research Institute, Beijing Jiaotong University, Weihai 264401, China}

\date{\today}

\begin{abstract}
We show that gravitons can thermalize much faster than the universe expands in an effective scalar-mediated model. A thermal fermion bath maintains a massive scalar close to equilibrium, while scalar decay and inverse decay, $\phi\rightleftarrows g+g$, drive the graviton distribution toward a Bose--Einstein distribution with zero chemical potential. We establish this equilibrium state and calculate how it is reached from an initially empty graviton sector. For a scalar mass of $1\,{\rm TeV}$, an interaction scale $\Lambda_{\rm int}=m_\phi/0.03$, and a Yukawa coupling $y=0.03$, the graviton occupations approach their equilibrium values to within $0.1\%$ over $0.1\leq p/T\leq5$ after approximately $0.024$ initial Hubble times, including cosmic expansion and energy transfer from the bath. This interval contains the thermal energy-spectrum peak. The scalar remains close to equilibrium, its vacuum decay widths are narrow, and the graviton interaction is perturbative over the reaction domain. Assuming the effective theory applies on this domain, our results demonstrate that gravitons can reach near-thermal equilibrium under these conditions. We also derive the present-day thermal spectrum for a specified decoupling and entropy history.
\end{abstract}

\maketitle

\section{Introduction}
\label{sec:introduction}

Primordial gravitational waves can carry information about the early universe if they propagate with little subsequent interaction~\cite{Caprini2018}. If gravitons instead exchange energy and particles efficiently with a thermal bath, their distribution can approach thermal equilibrium. Their spectrum would then reflect the bath temperature rather than the details of their production. This raises a basic question: can gravitons thermalize before cosmic expansion significantly changes the bath?

Producing gravitons from thermal matter is not sufficient to establish a thermal graviton distribution. A hot plasma can emit gravitational waves while the graviton population remains far from equilibrium~\cite{GhiglieriLaine2015,Ringwald2021,Ghiglieri2024}. Thermalization requires both emission and absorption to act efficiently. It also requires enough time for the distribution to approach equilibrium across the relevant momenta. To answer the question, one must therefore follow the graviton distribution rather than calculate only its production rate.

In this work, we show that rapid graviton thermalization is possible in an effective scalar-mediated heat-bath model. The model contains a thermal matter reservoir, a light Dirac fermion $\psi$, a massive real scalar $\phi$, and massless gravitons. A Yukawa interaction $y\phi\bar{\psi}\psi$ allows the bath to replenish the scalar population, while a scalar--Gauss--Bonnet coupling permits $\phi\rightleftarrows g+g$. Scalar couplings to quadratic curvature and their decays into gravitons have been studied as sources of high-frequency gravitational waves~\cite{EmaMukaidaNakayama2022,MudrunkaNakayama2024}. Here we include inverse decay as well as decay and determine whether these reactions bring the graviton population into equilibrium.

We consider a radiation-dominated epoch after the matter reservoir has formed. The scalar is a thermal mediator distinct from the inflaton, and no coherent scalar condensate is included. The reservoir maintains thermal fermion distributions with vanishing chemical potentials. If the Yukawa reactions are sufficiently rapid, they also keep the scalar close to equilibrium despite its decays into gravitons. The scalar can therefore continue to mediate energy exchange between the reservoir and the graviton sector.

We first identify the equilibrium state of the graviton collision equation. At fixed bath temperature, a relative-free-energy dissipation identity shows that collisions decrease the free energy and that the Bose--Einstein distribution with zero chemical potential is the unique regular stationary state on the connected reaction window. We then solve the nonlinear kinetic equation from an initially empty graviton sector to determine how quickly this state is approached.

For the parameter choice studied below, the graviton occupations approach their equilibrium values to within $0.1\%$ over $0.1\leq p/T\leq5$ after approximately $0.024$ initial Hubble times, including expansion and energy transfer from a finite reservoir. This momentum interval contains the thermal energy-spectrum peak. The bath keeps the scalar close to equilibrium throughout the calculated evolution, its vacuum decay widths remain narrow, and the graviton interaction remains perturbative over the reaction domain. Assuming the effective theory is valid on this domain, the calculation demonstrates that gravitons can reach near-thermal equilibrium well before expansion significantly changes the bath.

Our calculation establishes the formation of a near-thermal distribution from an empty graviton sector. It does not directly calculate the erasure of features in a pre-existing primordial spectrum, which would require evolving the corresponding initial distribution. Likewise, the final relic spectrum depends on how the gravitons decouple as the universe cools. Thermal graviton relics have also been considered in models with extra dimensions~\cite{SiegelFry2005}. Here we give the general mapping from a decoupled distribution to its present spectrum and derive the thermal limit under complete equilibration, instantaneous decoupling, and standard entropy evolution.

Section~\ref{sec:bath_thermalization} introduces the model, derives the kinetic equation, and determines its equilibrium state. Section~\ref{sec:thermalization_window} calculates the approach to equilibrium, first at fixed temperature and then with expansion and energy transfer included. Section~\ref{sec:thermalization_consistency} checks the conditions needed to maintain the scalar population and use the kinetic description. Section~\ref{sec:present_spectrum} discusses the present-day spectrum under specified decoupling conditions. We summarize the thermalization result in Sec.~\ref{sec:conclusion}.

\section{Interaction Mechanism and Thermal Equilibrium}
\label{sec:bath_thermalization}

We first explain how the bath exchanges energy with gravitons and determine the equilibrium distribution of the collision equation. The model contains a massive real scalar $\phi$, a light Dirac fermion $\psi$, a thermal matter reservoir, and the two helicities of the massless graviton. The scalar connects the reservoir to the graviton sector: bath reactions replenish the scalar, and scalar decay and inverse decay produce and absorb gravitons.

The calculation begins after reheating, during a radiation-dominated epoch with negligible further energy injection from the inflaton. The scalar is a thermal particle distinct from the inflaton, and no coherent scalar condensate is included. We assume isotropic distributions in a spatially flat Friedmann--Lema\^itre--Robertson--Walker universe. We use $\hbar=c=k_{\rm B}=1$ and the reduced Planck mass $M_{\rm Pl}=(8\pi G)^{-1/2}$.

The mediator interactions are specified by
\begin{equation}
    \mathcal{L}=\mathcal{L}_0-\frac{\phi}{\Lambda_{\rm GB}}\mathcal{G}-y\phi\bar{\psi}\psi+\mathcal{L}_{\rm bath},
    \label{eq:mediator_lagrangian}
\end{equation}
where $\mathcal{L}_0$ contains the Einstein--Hilbert term $M_{\rm Pl}^2R/2$ and the scalar terms $-(\nabla\phi)^2/2-m_\phi^2\phi^2/2$. The Gauss--Bonnet invariant is $\mathcal{G}=R_{\mu\nu\rho\sigma}R^{\mu\nu\rho\sigma}-4R_{\mu\nu}R^{\mu\nu}+R^2$. The fermion is vectorlike under any bath gauge symmetries, and $\phi$ is a singlet under them, so that $y\phi\bar\psi\psi$ is gauge invariant. The term $\mathcal{L}_{\rm bath}$ includes the fermion kinetic term, the remaining reservoir fields, and their interactions. We take $m_\psi\ll m_\phi$ and neglect fermion masses in the collision integrals.

The reservoir maintains the fermions at temperature $T(t)$ with vanishing chemical potentials, so that $f_\psi(p,t)=f_{\bar\psi}(p,t)=[e^{p/T(t)}+1]^{-1}$. We assume that its internal interactions restore these distributions faster than expansion and scalar reactions disturb them. We also assume that the fermion dispersion relations remain approximately massless. The microscopic bath interactions are not specified. Below, $g_*=106.75$ is used as a reference value for its energy density.

The Yukawa interaction allows $\psi+\bar\psi\rightleftarrows\phi$ and keeps the scalar close to equilibrium when these reactions are sufficiently rapid. The curvature interaction allows $\phi\rightleftarrows g+g$. Scalar decay transfers energy to gravitons, while inverse decay transfers it back. Individual scalars can therefore decay without depleting the thermal scalar population, because the bath continually replaces them. These reactions also change graviton number.

Initially, the reservoir and scalar are in equilibrium at $T_i=m_\phi$, and the graviton sector is empty. Both graviton helicities have the same distribution. We first hold the bath temperature fixed and then allow it to change through expansion and energy transfer. In the numerical calculation, we restrict reactions to scalar energies below $E_{\rm cut}$ in the bath frame and assume that the effective interaction is valid over this domain.

Defining $\Lambda_{\rm int}^3\equiv\Lambda_{\rm GB}M_{\rm Pl}^2$, the vacuum decay width into two gravitons is~\cite{EmaMukaidaNakayama2022,MudrunkaNakayama2024}
\begin{equation}
    \Gamma_g
    =\frac{m_\phi^7}{4\pi\Lambda_{\rm int}^6}.
    \label{eq:mediator_graviton_width}
\end{equation}
The scale $\Lambda_{\rm int}$ sets the strength of the graviton vertex; it is not necessarily the cutoff of the complete effective theory. For one light Dirac fermion, the vacuum width into fermions is $\Gamma_b=(y^2m_\phi/8\pi)(1-4m_\psi^2/m_\phi^2)^{3/2}$. These widths alone do not determine thermalization. The rates in a populated bath also depend on inverse reactions and Bose or Fermi statistical factors~\cite{Weldon1983}. We include these effects in the collision equations and estimate the leading thermal-mass correction in Sec.~\ref{sec:thermalization_consistency}.

Let $f_g(p,t)$ be the occupation number per graviton polarization at physical momentum $p$. When the scalar is in equilibrium, its distribution is $F_\phi^{\rm eq}(E,t)=[e^{E/T(t)}-1]^{-1}$. With $f_p=f_g(p,t)$ and $f_q=f_g(q,t)$, the difference between emission and absorption is~\cite{UehlingUhlenbeck1933}
\begin{equation}
    \mathcal{B}(p,q)=F_\phi^{\rm eq}(p+q,t)(1+f_p+f_q)-f_pf_q.
\end{equation}
Here emission contributes $F_\phi^{\rm eq}(1+f_p)(1+f_q)$ and absorption contributes $(1+F_\phi^{\rm eq})f_pf_q$. Their difference gives the expression above. Energy--momentum conservation requires $m_\phi^2=2pq(1-\cos\theta)$, so a pair can react only if $pq\geq m_\phi^2/4$.

Scalar decay first fills the empty graviton sector. As the occupations grow, stimulated emission and inverse decay both become important. For a narrow scalar quasiparticle, neglecting thermal changes to the matrix element and dispersion relations, the distribution obeys
\begin{equation}
\begin{split}
(\partial_t-Hp\partial_p)f_g(p,t) =\frac{m_\phi\Gamma_g}{p^2} \int_{m_\phi^2/(4p)}^\infty dq\,\mathcal{B}(p,q).
\end{split}
\label{eq:graviton_collision}
\end{equation}
The left-hand side accounts for time evolution and the redshift of physical momentum. We denote the collision term on the right-hand side by $C_g[p;f_g,F_\phi^{\rm eq}]$. Any restriction on the allowed reactions is applied equally to emission and absorption, so that detailed balance is preserved.

The Bose--Einstein distribution
\begin{equation}
    f_g^{\rm eq}(p,t)=\frac{1}{e^{p/T(t)}-1}
    \label{eq:graviton_equilibrium}
\end{equation}
makes the collision term vanish. Chemical equilibrium requires $\mu_\phi=2\mu_g$. Since the bath maintains $\mu_\phi=0$, the equilibrium graviton chemical potential is also zero. We next show that this is the unique regular stationary distribution on the reaction window and that collisions decrease the relative free energy.

For this argument, hold the temperature fixed and neglect expansion. Let $I$ be the connected interval of participating momenta, and let $\mathcal D$ contain the allowed pairs satisfying $pq\geq m_\phi^2/4$ and $p+q<E_{\rm cut}$. This domain is symmetric under $p\leftrightarrow q$. Write $n_p=(e^{p/T}-1)^{-1}$. The relative free energy divided by $T$, including both helicities, is
\begin{equation}
\begin{split}
    \mathcal H[f]=\frac{1}{\pi^2}\int_I dp\,p^2
    \Biggl[&f_p\ln\frac{f_p}{n_p} -(1+f_p)\ln\frac{1+f_p}{1+n_p}\Biggr].
\end{split}
\label{eq:relative_free_energy}
\end{equation}
The integrand is nonnegative and vanishes only at $f_p=n_p$. Define the gain factor $A_{pq}=F_\phi^{\rm eq}(p+q)(1+f_p)(1+f_q)$ and the loss factor $B_{pq}=[1+F_\phi^{\rm eq}(p+q)]f_pf_q$. Substituting Eq.~\eqref{eq:graviton_collision} into the time derivative and symmetrizing under $p\leftrightarrow q$ gives
\begin{equation}
    \frac{d\mathcal H}{dt}
    =-\frac{m_\phi\Gamma_g}{2\pi^2}
    \int_{\mathcal D}dp\,dq\,
    (A_{pq}-B_{pq})\ln\frac{A_{pq}}{B_{pq}}.
    \label{eq:free_energy_dissipation}
\end{equation}
For positive occupations, the integrand is nonnegative, so collisions decrease the relative free energy. An initially empty distribution is treated by the continuous limit of Eq.~\eqref{eq:relative_free_energy}, and the dissipation identity applies after occupations become positive in the interior of $I$.

For a regular stationary distribution, the dissipation rate vanishes and therefore $A_{pq}=B_{pq}$ throughout the reaction domain. Define $\chi(p)=\ln[f_p/(1+f_p)]+p/T$. Detailed balance then gives $\chi(p)+\chi(q)=0$ for each allowed pair. Nearby interior momenta share allowed partners, so they have the same value of $\chi$. Since $I$ is connected, $\chi$ is constant on $I$, and the pair condition forces that constant to be zero. Thus $f_p=n_p$ is the unique regular stationary distribution. The free-energy identity establishes its stability, but does not give a relaxation time, especially near the edges of the reaction window.

The numerical calculation below tests the formation of this distribution from an empty graviton sector. Erasure of a particular pre-existing spectral feature would require evolving that initial distribution. Occupation numbers also do not specify the evolution of quantum coherence or squeezing.

To allow for a finite reservoir, we must also track the energy removed from the bath. If both graviton helicities equilibrate, their energy density is $\rho_g^{\rm eq}=\pi^2T^4/15$. Producing this population can change the bath temperature.

Let $\rho_{\rm bath}$ and $P_{\rm bath}$ include the matter plasma and the mediator when it remains in thermal equilibrium. The net energy-transfer rate into gravitons is
\begin{equation}
    Q_g=\frac{1}{\pi^2}\int_0^\infty dp\,p^3
    C_g[p;f_g,F_\phi^{\rm eq}],
    \label{eq:graviton_energy_transfer}
\end{equation}
where the prefactor includes both graviton polarizations. The coupled energy equations are
\begin{equation}
    \dot{\rho}_g+4H\rho_g=Q_g,
    \label{eq:graviton_energy_balance}
\end{equation}
\begin{equation}
    \dot{\rho}_{\rm bath}
    +3H(\rho_{\rm bath}+P_{\rm bath})=Q_{\rm in}-Q_g,
    \label{eq:coupled_energy_budget}
\end{equation}
where $Q_{\rm in}$ denotes any remaining energy injection from the inflaton sector. It can be neglected during the radiation-dominated interval considered below.

For a reference bath density $\rho_{\rm bath}=(\pi^2/30)g_*T^4$, the equilibrium ratio is $\rho_g^{\rm eq}/\rho_{\rm bath}=2/g_*$. Here $g_*$ counts the effective energy degrees of freedom included in the bath. A large reservoir therefore loses only a small fraction of its energy to equilibrated gravitons. This motivates the fixed-temperature approximation over a short interval; the expanding calculation instead evolves the temperature using the coupled energy equations.

The result of this section is that the scalar reactions select a Bose--Einstein distribution with zero chemical potential as the unique regular stationary state on the connected reaction window. At fixed temperature, collisions decrease the relative free energy. The remaining question is whether the distribution approaches equilibrium quickly enough in an expanding universe.

\section{Thermalization within a Hubble Time}
\label{sec:thermalization_window}

We now calculate how long gravitons take to approach equilibrium and compare this time with the expansion timescale. We solve the nonlinear kinetic equation from an empty graviton sector, using a scalar in equilibrium with the bath. We first keep the temperature fixed, then include expansion and the energy drawn from a finite reservoir. In both calculations, we test the occupations over $0.1\leq p/T\leq5$, which contains the thermal energy-spectrum peak.

For a radiation bath with approximately constant effective degrees of freedom, the reference expansion rate is
\begin{equation}
    H(T)=\sqrt{\frac{\pi^2g_*}{90}}\frac{T^2}{M_{\rm Pl}}.
\end{equation}
We define $H_m=H(T=m_\phi)$ and use $H_m^{-1}$ as the reference Hubble time. The expanding calculation below includes the energy transferred from the bath to gravitons.

We use $m_\phi=10^3\,{\rm GeV}$, $m_\phi/\Lambda_{\rm int}=0.03$, and $y=0.03$ throughout the kinetic calculation and consistency analysis. We take $g_*=106.75$ and $M_{\rm Pl}=2.435\times10^{18}\,{\rm GeV}$. The bath count is a fixed reference value; corrections from additional thermal species are neglected. Equation~\eqref{eq:mediator_graviton_width} gives $\Gamma_g\simeq5.80\times10^{-8}\,{\rm GeV}$, while $H_m\simeq1.41\times10^{-12}\,{\rm GeV}$. We use $\Gamma_g^{-1}$ as the time unit and determine equilibration from the evolving distribution.

A graviton with momentum $p$ needs a partner with $q\geq m_\phi^2/(4p)$ to form an on-shell scalar. The restriction $p+q<E_{\rm cut}$ also requires $q<E_{\rm cut}-p$. If these bounds leave no interval, the collision term is zero. Very soft gravitons require energetic partners, so the scalar-energy restriction excludes both very soft modes and modes too energetic to leave room for a partner. This bath-frame restriction defines the computational domain; it does not establish the effective-theory cutoff.

The condition $p+m_\phi^2/(4p)<E_{\rm cut}$ restricts the momentum to $p_-<p<p_+$, where
\begin{equation}
    p_\pm=\frac{E_{\rm cut}\pm\sqrt{E_{\rm cut}^2-m_\phi^2}}{2}.
    \label{eq:reaction_window}
\end{equation}
We choose $E_{\rm cut}=8m_\phi$. At $T=m_\phi$, the corresponding interval in $x\equiv p/T$ is $0.031373<x<7.968627$. Modes outside this interval do not interact in the fixed-temperature calculation. Cosmological cooling changes the location of the window in $x$.

We take an initially empty graviton sector, $f_g(x,0)=0$, hold the temperature fixed at $T=m_\phi$, and introduce $u=\Gamma_gt$. Writing $f_x\equiv f_g(x,u)$ and $n(z)\equiv(e^z-1)^{-1}$, we define the dimensionless gain-minus-loss term as $\mathcal B_{xv}=n(x+v)(1+f_x+f_v)-f_xf_v$. Neglecting expansion, Eq.~\eqref{eq:graviton_collision} becomes
\begin{equation}
    \partial_u f_x=\frac{1}{x^2}\int_{1/(4x)}^{8-x}dv\,\mathcal B_{xv}.
    \label{eq:nonlinear_dimensionless}
\end{equation}
Both occupations in $\mathcal B_{xv}$ are evaluated at the same time $u$. The collision term is set to zero when the lower integration limit is greater than or equal to the upper limit. Detailed balance gives $\mathcal B_{xv}=0$ for $f_x=n(x)$, so the Bose--Einstein distribution is stationary. Modes outside the reaction window retain their initial occupation.

We discretize the collision integral using Gaussian quadrature in $\ln x$ over the reaction window and evolve the resulting equations with an implicit time integrator. The gain and loss terms use the same reaction domain and preserve the equilibrium solution. Calculations with $320$, $640$, and $960$ quadrature nodes support the rounded results below.

To measure the departure from equilibrium, we define
\begin{equation}
    \mathcal{E}_{0.1,5}(u)
    \equiv\max_{0.1\leq x\leq5}
    \left|\frac{f_g(x,u)}{n(x)}-1\right|.
    \label{eq:distribution_error}
\end{equation}
The tested interval lies inside the reaction window and contains the equilibrium energy-spectrum peak per logarithmic momentum interval, $x_{\rm peak}\simeq3.92069$. Starting from an empty graviton sector at fixed temperature $T=m_\phi$, the nonlinear evolution gives $\mathcal{E}_{0.1,5}\sim10^2$ at $\Gamma_gt=10$, $\mathcal{E}_{0.1,5}\sim0.2$ at $\Gamma_gt=100$, and $\mathcal{E}_{0.1,5}<10^{-4}$ at $\Gamma_gt=1000$.

The large error at early times comes from a temporary excess of low-momentum gravitons. Some modes rise above their equilibrium occupations before emission and absorption balance. This is consistent with the decreasing relative free energy, which constrains the distribution as a whole rather than requiring each occupation to change monotonically. By $u=1000$, the excess has subsided and the relative error is below $10^{-4}$ across the tested band.

Defining the sampled equilibration time as $t_*\equiv1000/\Gamma_g$, we obtain
\begin{equation}
    H_mt_*\simeq2.42\times10^{-2}.
    \label{eq:local_equilibration_time}
\end{equation}
At this sampled time, both helicities satisfy the stated criterion; $t_*$ is not a determination of the earliest time at which they do so. Since it is much shorter than $H_m^{-1}$, we next test whether equilibration survives expansion and the change in bath temperature.

At fixed $T/m_\phi=1$ and $E_{\rm cut}/m_\phi=8$, the dimensionless collision equation is independent of the overall decay width. The same sampled criterion at $u_*=1000$ is reached within one reference Hubble time if $m_\phi/\Lambda_{\rm int}>0.0161$ for $m_\phi=1\,{\rm TeV}$ and $g_*=106.75$. The chosen ratio $0.03$ satisfies this condition in the fixed-temperature calculation.

We next include cosmological cooling and energy transfer to gravitons by evolving the same collision model in an expanding, finite reservoir. We take $\rho_{\rm bath}=(\pi^2/30)g_*T^4$, $P_{\rm bath}=\rho_{\rm bath}/3$, constant $g_*=106.75$, and $Q_{\rm in}=0$. The temperature-dependent contribution of the massive mediator to this radiation equation of state is neglected. We choose the initial time as $t=0$, normalize $a(0)=1$, and set $T(0)=m_\phi$. We retain $u=\Gamma_gt$ and introduce the dimensionless comoving momentum $k=ap/m_\phi$. Let $\mathfrak f(k,u)=f_g(m_\phi k/a,t)$.

For this equation of state, total comoving radiation energy is conserved. The scale factor is $a(u)=[1+2(H_m/\Gamma_g)u]^{1/2}$, and the bath temperature is determined by
\begin{equation}
    \left(\frac{aT}{m_\phi}\right)^4
    =1-\frac{30}{\pi^4g_*}
    \int_0^\infty dk\,k^3\mathfrak f(k,u).
    \label{eq:finite_bath_temperature}
\end{equation}
Expansion reduces the temperature, and graviton production removes additional energy from the bath. The latter effect appears as a decrease in $aT$ in Eq.~\eqref{eq:finite_bath_temperature}. In the radiation equation of state used here, energy exchange changes the division of energy between the two sectors but not their total comoving energy.

We evolve the graviton equation in comoving momentum and update $T$ from Eq.~\eqref{eq:finite_bath_temperature}. The scalar follows its instantaneous equilibrium distribution. Allowed partner momenta satisfy $k\ell\geq a^2/4$ and $k+\ell<8a$, so the reaction domain moves as the universe expands. Emission and absorption use this same symmetric domain. We use Gaussian quadrature in $\ln k$ and an implicit integrator, with a grid covering the full reaction window during the calculation.

We compare the expanding solution with the Bose--Einstein distribution at the instantaneous bath temperature. At $u=1000$, the graviton occupation differs from its equilibrium value by less than $0.1\%$ at every momentum in the interval $0.1\leq p/T(t_*)\leq5$. Refinement of the comoving grid from $240$ to $480$ and $960$ nodes supports this result.

At $t_*=1000/\Gamma_g$, the bath temperature is approximately $0.9723m_\phi$. The distribution therefore stays close to equilibrium even while the bath cools. The error thresholds refer to the solutions of the adopted kinetic equation; they do not include omitted thermal or higher-order corrections.

As a separate check at fixed temperature, calculations with 240 momentum nodes and $E_{\rm cut}/m_\phi=6,8,12,16$ all give $\mathcal E_{0.1,5}(1000)<10^{-4}$.

The tested band does not include all momenta. Modes outside the fixed-temperature reaction window remain empty, while modes can enter or leave the window in the expanding calculation. Following this process through decoupling is necessary to obtain the complete relic spectrum.

The main result is that, for the stated parameters, gravitons reach near-thermal equilibrium over $0.1\leq p/T\leq5$ within approximately $0.024$ initial Hubble times. The maximum relative occupation error is below $10^{-4}$ at fixed temperature and below $10^{-3}$ with expansion and energy transfer included. Thus, within this calculation, expansion does not prevent thermalization over the momentum interval containing the thermal energy-spectrum peak.

\section{Conditions for the Thermalization Calculation}
\label{sec:thermalization_consistency}

We now check whether the conditions assumed in the kinetic calculation can hold during thermalization. The bath must keep the scalar close to equilibrium despite graviton production, and the scalar must remain narrow enough for the decay description to apply. The graviton vertex must also remain perturbative. We use the same parameters as in Sec.~\ref{sec:thermalization_window}, beginning at $T=m_\phi$.

For one light Dirac fermion, the fermionic decay width given above yields $\Gamma_b\simeq3.58\times10^{-2}\,{\rm GeV}$. The vacuum channels included here give $(\Gamma_b+\Gamma_g)/m_\phi\simeq3.58\times10^{-5}$, consistent with the narrow-width approximation. Additional thermal contributions to the width must also remain small compared with $m_\phi$.

The dimensionless strength of the canonically normalized scalar--graviton vertex at characteristic energy $E$ scales as
\begin{equation}
    g_{\rm eff}(E)\sim\left(\frac{E}{\Lambda_{\rm int}}\right)^3.
\end{equation}
At $E_{\rm cut}=8m_\phi$, this gives $g_{\rm eff}\simeq1.38\times10^{-2}$, so the vertex is perturbative.

The next question is whether the bath can replace scalars as rapidly as graviton reactions disturb their distribution. A small vacuum graviton width alone does not answer this, because large graviton occupations enhance the reactions during the transient stage. We therefore evaluate the disturbance using the evolving graviton distribution.

For a scalar of energy $E$ and momentum $P=\sqrt{E^2-m_\phi^2}$, the decay-product energies are $p=(E+P\mu)/2$ and $q=(E-P\mu)/2$. Here $\mu$ is the rest-frame decay-angle cosine relative to the boost direction. Define the isotropic average $\langle X\rangle_\mu=\frac12\int_{-1}^{1}d\mu\,X$. With $f_p=f_g(p,t)$ and $f_q=f_g(q,t)$, the graviton contribution to the scalar collision term is
\begin{equation}
\begin{split}
    C_{\phi g}(E,t)=\frac{m_\phi\Gamma_g}{E}
    \Bigl[ \langle f_pf_q\rangle_\mu -F_\phi\langle1+f_p+f_q\rangle_\mu\Bigr].
\end{split}
\label{eq:scalar_graviton_collision}
\end{equation}
Here $F_\phi=F_\phi(E,t)$ denotes the actual scalar distribution. This contribution is included only for $E<E_{\rm cut}$, consistently with the graviton calculation.

For massless equilibrium fermions with $n_F(p)=(e^{p/T}+1)^{-1}$, the bath collision term is $C_{\phi b}=-\gamma_{\phi b}(E)[F_\phi-F_\phi^{\rm eq}]$, where
\begin{equation}
    \gamma_{\phi b}(E)
    =\frac{m_\phi\Gamma_b}{E}
    \left\langle1-n_F(p)-n_F(q)\right\rangle_\mu.
    \label{eq:boosted_scalar_bath_rate}
\end{equation}
This expression includes time dilation and Pauli blocking, while neglecting fermion thermal masses. The combination of decay and inverse-decay contributions follows the finite-temperature relaxation-rate structure discussed in Ref.~\cite{Weldon1983}.

We measure the disturbance relative to the bath response by
\begin{equation}
    \epsilon_\phi(E,t)
    \equiv
    \frac{|C_{\phi g}[F_\phi^{\rm eq},f_g](E,t)|}
    {\gamma_{\phi b}(E)F_\phi^{\rm eq}(E)}.
    \label{eq:mediator_departure_diagnostic}
\end{equation}
For a prescribed graviton distribution at fixed temperature, set $\delta F_\phi=F_\phi-F_\phi^{\rm eq}$ and define $\lambda_\phi(E,t)=\gamma_{\phi b}(E)+(m_\phi\Gamma_g/E)\langle1+f_p+f_q\rangle_\mu$. The scalar equation is linear in its departure from equilibrium:
\begin{equation}
    \partial_t\delta F_\phi+\lambda_\phi\delta F_\phi
    =C_{\phi g}[F_\phi^{\rm eq},f_g].
    \label{eq:mediator_departure_evolution}
\end{equation}
The graviton reactions also contribute to restoring the scalar distribution, so $\lambda_\phi\geq\gamma_{\phi b}$. For an initially equilibrated scalar at fixed temperature, the integrating-factor solution bounds its fractional departure at each energy by the largest preceding value of $\epsilon_\phi(E,t)$. A small $\epsilon_\phi$ therefore means that the bath can keep the scalar close to equilibrium even during graviton production. The sampled values below estimate this bound.

During the nonlinear evolution, the maximum sampled value of $\epsilon_\phi$ is approximately $3\times10^{-5}$ over $m_\phi\leq E<E_{\rm cut}$ and $0\leq\Gamma_gt\leq1000$. This includes the temporary excess of gravitons and is stable at the quoted order when the momentum grid is refined from $640$ to $960$ nodes. The bath restoring rate also greatly exceeds $H_m$ throughout this energy interval.

In an expanding universe, the scalar must also follow the changing bath temperature. Its departure equation then has an additional source, $-(\partial_t-HP\partial_P)F_\phi^{\rm eq}$. In the $960$-node calculation, the sampled value of $\epsilon_\phi$ remains below $3\times10^{-5}$. The additional source, divided by $\gamma_{\phi b}F_\phi^{\rm eq}$, remains below $6\times10^{-7}$. Thus the bath response is fast compared with both graviton-induced disturbances and the change in the scalar equilibrium distribution.

Using the leading high-temperature Yukawa estimate $\delta m_\phi^2\simeq y^2T^2/6$~\cite{HidakaSatowKunihiro2012}, we obtain $\delta m_\phi^2/m_\phi^2\simeq1.5\times10^{-4}$ at $T=m_\phi$. This estimate suggests a small thermal-mass correction, which we neglect in the collision calculation. The fermion's other bath interactions must also leave the $\phi\leftrightarrow\bar{\psi}\psi$ channel open. Checking their thermal-mass and damping effects requires specifying those interactions.

Complete equilibration of both graviton helicities would give an energy density approximately $1.9\%$ of the reference bath energy density. Bath backreaction is therefore small, although its effect on the temperature is included in the expanding calculation.

Perturbativity of the vertex does not establish the cutoff of the complete effective theory or control independent higher-order operators. Causality bounds for a massless scalar illustrate this distinction~\cite{Serra2022}. Their application to the massive mediator with Yukawa couplings requires a separate analysis. We therefore retain the assumption that the effective curvature interaction is valid over the reaction domain.

For the chosen parameters, the checks show that the bath can keep the scalar close to equilibrium during the calculated evolution. The vacuum widths are narrow, the graviton vertex is perturbative, and the estimated Yukawa thermal-mass correction is small. These results support the approximations used to obtain rapid graviton thermalization, subject to the stated assumptions about the bath and effective theory.

\section{Present-Day Spectrum in the Thermal Limit}
\label{sec:present_spectrum}

We now determine how a distribution left after decoupling appears today. First we give the redshift relation for an arbitrary graviton distribution. We then derive the thermal spectrum assuming complete equilibration, instantaneous decoupling, and standard entropy evolution. This thermal limit extends beyond the finite momentum band and short evolution interval tested above; it is not yet the relic prediction of the mediator calculation.

Let $t_f$ denote a time after which graviton interactions are negligible over the momentum range of interest, and write $a_f=a(t_f)$. Collisionless propagation preserves the occupation number along a freely redshifting trajectory:
\begin{equation}
    f_{g,0}(p_0)
    =f_g\!\left(p_0\frac{a_0}{a_f},t_f\right).
    \label{eq:collisionless_distribution}
\end{equation}
For an isotropic, unpolarized population, the energy density is $\rho_g=\pi^{-2}\int_0^\infty dp\,p^3f_g(p)$, including both graviton polarizations. Defining $\Omega_{g,0}(f)\equiv\rho_{c,0}^{-1}d\rho_{g,0}/d\ln f$ and $\rho_{c,0}=3M_{\rm Pl}^2H_0^2$, we obtain
\begin{equation}
    \Omega_{g,0}(f)
    =\frac{p_0^4}{\pi^2\rho_{c,0}}
    f_g\!\left(p_0\frac{a_0}{a_f},t_f\right).
    \label{eq:general_redshifted_distribution}
\end{equation}
Momentum and energy density in this equation are in natural units. For frequency $f$ in Hz, use $p_0=2\pi\hbar f$ with $c=1$. This mapping applies to both thermal and nonthermal distributions.

Suppose that the graviton distribution is Planckian at decoupling, with negligible chemical potential and $T_{g,\rm dec}=T_{\rm dec}$. Under instantaneous decoupling and subsequent free propagation, its temperature parameter redshifts as $T_{g,0}=T_{\rm dec}a_{\rm dec}/a_0$. Assuming standard entropy evolution in the remaining matter sector~\cite{SaikawaShirai2018} gives
\begin{equation}
    T_{g,0}
    =T_{\gamma,0}
    \left(\frac{g_{*s,0}}{g_{*s,\rm dec}}\right)^{1/3},
    \label{eq:graviton_temperature_today}
\end{equation}
Here $g_{*s,\rm dec}$ counts the bath entropy degrees of freedom, excluding the decoupled gravitons. As bath species disappear from the thermal population, their entropy heats the remaining matter relative to the freely propagating gravitons. The graviton temperature today is therefore lower than the photon temperature. We assume no additional entropy injection beyond this thermal history.

Taking $T_{\gamma,0}=2.725\,{\rm K}$~\cite{Fixsen2009} and $g_{*s,0}=3.91$ gives $T_{g,0}\simeq0.905\,{\rm K}\,(106.75/g_{*s,\rm dec})^{1/3}$. The reference value $g_{*s,\rm dec}=106.75$ specifies the entropy history for this example. It does not determine when the mediator model decouples and need not equal the energy count $g_*$ used earlier.

For two massless graviton polarizations with a Planck distribution, the energy density per logarithmic momentum interval is
\begin{equation}
    \frac{d\rho_g}{d\ln p}
    =\frac{1}{\pi^2}\frac{p^4}{e^{p/T_g}-1}.
    \label{eq:thermal_energy_spectrum}
\end{equation}
Integration gives $\rho_g=\pi^2T_g^4/15$. For physical frequency $f$ in Hz, define $x\equiv2\pi\hbar f/(k_{\rm B}T_{g,0})$. The present-day spectrum is
\begin{equation}
    \Omega_{g,0}(f)
    =\frac{15}{\pi^4}\Omega_{\gamma,0}
    \left(\frac{g_{*s,0}}{g_{*s,\rm dec}}\right)^{4/3}
    \frac{x^4}{e^x-1}.
    \label{eq:normalized_thermal_spectrum}
\end{equation}
Here $\Omega_{\gamma,0}$ is the present photon energy-density parameter, and the factor $15/\pi^4$ normalizes the Planck spectral shape. The equilibrium spectrum scales as $f^3$ for $x\ll1$ and as $f^4e^{-x}$ for $x\gg1$.

The maximum of $x^4/(e^x-1)$ satisfies $4(1-e^{-x_{\rm peak}})=x_{\rm peak}$, giving $x_{\rm peak}\simeq3.92069$. Thus $f_{\rm peak}=x_{\rm peak}k_{\rm B}T_{g,0}/(2\pi\hbar)$, or
\begin{equation}
    f_{\rm peak}\simeq7.39\times10^{10}\,{\rm Hz}
    \left(\frac{106.75}{g_{*s,\rm dec}}\right)^{1/3}.
    \label{eq:thermal_peak_frequency_today}
\end{equation}
For $H_0=70\,{\rm km\,s^{-1}\,Mpc^{-1}}$, the adopted photon temperature gives $\Omega_{\gamma,0}\simeq5.05\times10^{-5}$. Evaluating Eq.~\eqref{eq:normalized_thermal_spectrum} at $x_{\rm peak}$ gives
\begin{equation}
    \Omega_{g,0}^{\rm peak}\simeq4.5\times10^{-7}
    \left(\frac{106.75}{g_{*s,\rm dec}}\right)^{4/3}.
    \label{eq:thermal_peak_amplitude_today}
\end{equation}
Once $g_{*s,\rm dec}$ is specified, the thermal limit fixes the spectral shape, peak frequency, and amplitude together. These properties probe the assumed equilibrium and entropy history. A thermal shape alone would not identify the scalar mediator, since other mechanisms could produce the same distribution.

The peak lies in the millimetre-wave band, motivating searches based on graviton--photon conversion in a strong magnetic field. Dielectric haloscopes of the MADMAX type provide one possible approach: a suitably arranged stack of dielectric disks enhances the electromagnetic signal generated by an incident gravitational wave. Configurations optimized for gravitational waves have been studied in the tens-of-GHz range, relevant to the frequency band of our thermal prediction~\cite{DomckeEllisKopp2025}. Testing the Planck shape would require measurements at several frequencies, rather than an excess in a single narrow band.

The projected sensitivity in Ref.~\cite{DomckeEllisKopp2025} is insufficient for this thermal benchmark. The spectrum is therefore a target for future detector development. A quantitative assessment would need the response to an isotropic stochastic background, together with detector noise, bandwidth, and observation time.

For the mediator model itself, the coupled evolution must be continued through momentum-dependent decoupling to determine $f_g(p,t_f)$ and $a_f/a_0$. Equation~\eqref{eq:general_redshifted_distribution} then gives the relic spectrum, including any deviations caused by incomplete equilibration or gradual decoupling.

The result of this section is a general relation between the decoupled distribution and the present spectrum, together with its complete-equilibrium limit. With instantaneous decoupling, standard entropy evolution, and $g_{*s,\rm dec}=106.75$, that limit gives $T_{g,0}\simeq0.905\,{\rm K}$, a peak near $74\,{\rm GHz}$, and $\Omega_{g,0}^{\rm peak}\simeq4.5\times10^{-7}$ for the adopted $H_0$. These are conditional thermal-relic values; the preceding sections establish the separate dynamical result that gravitons can approach equilibrium rapidly over the tested momentum interval.

\section{Conclusion}
\label{sec:conclusion}

Gravitons can reach near-thermal equilibrium in the scalar-mediated model studied here. The thermal bath continually replenishes the scalar population, while scalar decay produces gravitons and inverse decay absorbs them. Together, these reactions drive the graviton distribution toward thermal equilibrium. At fixed bath temperature, the Bose--Einstein distribution with zero chemical potential is the unique regular stationary state on the connected reaction window, and collisions decrease the relative free energy.

Our numerical calculation shows that this process can be fast enough to operate in an expanding universe. For $m_\phi=1\,{\rm TeV}$, $m_\phi/\Lambda_{\rm int}=0.03$, and $y=0.03$, an initially empty graviton sector approaches equilibrium over $0.1\leq p/T\leq5$ within approximately $0.024$ initial Hubble times. The maximum relative occupation error is below $10^{-4}$ at fixed temperature and below $10^{-3}$ when expansion and energy transfer from the bath are included. Thus, expansion does not prevent thermalization over this momentum interval, which contains the thermal energy-spectrum peak.

The bath keeps the scalar close to equilibrium throughout the calculated evolution. Its vacuum decay widths remain narrow, and the graviton interaction remains perturbative over the reaction domain. Assuming the effective theory is valid on this domain, these results provide a concrete example in which gravitons thermalize well within a Hubble time. The conclusion applies to the tested momentum interval; complete equilibration at all momenta has not been established.

If the graviton population reaches full thermal equilibrium before decoupling, its present spectrum is determined by the subsequent entropy history. Instantaneous decoupling with $g_{*s,\rm dec}=106.75$ gives a present temperature of approximately $0.905\,{\rm K}$ and a spectral peak near $74\,{\rm GHz}$. Obtaining the relic spectrum of the specific mediator model requires extending the evolution through decoupling. The result established here is that, under the stated conditions, scalar decay and inverse decay can bring gravitons into near-thermal equilibrium before cosmic expansion significantly changes the bath.

\section*{Acknowledgements}  
This work is supported by the Talent Fund of the Weihai Research Institute, Beijing Jiaotong University (2025WHRC001).

\end{document}